\documentclass[aps, prd, superscriptaddress, nofootinbib, preprintnumbers,twocolumn, showpacs]{revtex4}
\usepackage{graphicx}
\usepackage{amsmath,amssymb,array,calc,epsfig}
\usepackage{psfrag,verbatim,bm}
\preprint{UWO-TH-07/15}

\newcommand{\qq}{\mathfrak{q}}
\newcommand{\ww}{\mathfrak{w}}

\begin{document}

\def\call{{\cal{L}}}
\def\dd{{\delta}}
\def\T{{\tau}}
\def\del{{\partial}}
\def\e{{\epsilon}}
\def\ie{{\it i.e.}}
\def\ga{{\gamma}}
\def\caln{{\cal{N}}}

\title{Bulk viscosity of gauge theory plasma at strong coupling}

\author{Alex Buchel}
\affiliation{Perimeter Institute for Theoretical Physics, 
Waterloo, Ontario N2J 2W9, Canada}
\affiliation{
Department of Applied Mathematics, University of Western
Ontario, London, Ontario N6A 5B7, Canada
}

\date{August 25, 2007}

\begin{abstract}
We propose a lower bound on bulk viscosity of strongly coupled gauge
theory plasmas.  Using explicit example of the $\caln=2^*$ gauge
theory plasma we show that the bulk viscosity remains finite at a
critical point with a divergent specific heat. We present an estimate
for the bulk viscosity of QGP plasma at RHIC.
\end{abstract}


\maketitle

Recently, a holographic link between finite temperature  gauge theories and string theory
black holes emerged as a viable theoretical tool to model properties of strongly coupled 
quark gluon plasma (QGP) produced at RHIC \cite{rhic1,rhic2,rhic3,rhic4}. While the precise holographic dual to QCD is still missing, 
a progress in study of string theory black holes made it possible to compare 
the thermodynamics of strongly coupled QCD-like gauge theories \cite{bdkl,abk}
with lattice results \cite{kl}. The dual holographic approach has been successful to 
address dynamical properties of  QGP such as the shear viscosity \cite{pss} and 
the parton jet quenching \cite{lrw,hkkky}, where few alternative techniques are available. 
Intriguingly, dual string theory studies reveal 
certain universal features of gauge theory plasma dynamics. A notable examples is the 
ratio of the shear viscosity $\eta$ to the entropy density $s$. It was shown in 
\cite{u1,u2,u3,u4} that
\begin{equation}
\frac{\eta}{s}=\frac {1}{4\pi}\ \longrightarrow\ \frac{\hbar}{4\pi k_B}\approx 6.08\times 10^{-13}\ {\rm K\ s}\,,
\label{eta}
\end{equation}  
in any gauge theory plasma at infinite 't Hooft coupling, irrespectively of the  dimensionality of the space, 
the microscopic scales of the theory, and chemical potentials for the conserved quantities.  
The universality of the shear viscosity ratio (\ref{eta}) in  strongly coupled gauge theories at finite 
temperature  led Kovtun, Son and  Starinets (KSS) to conjecture a shear viscosity bound \cite{kss} 
\begin{equation}
\frac{\eta}{s}\ge \frac{1}{4\pi}\,,
\label{eta1}
\end{equation}
for all physical systems in Nature. Empirically, the KSS bound indeed appears to be satisfied by all common substances 
\cite{u2}; moreover, it is correct at large (but finite) 't Hooft coupling in $\caln=4$ Yang-Mills theory plasma \cite{cor1,cor2}.

We believe that it is such universal features of  dual holographic models of gauge theories that might have some relevance to 
QCD. Thus, it is imperative to ask what are other generic properties of strongly coupled gauge theories.   
The question is complicated as neither the bulk viscosity \cite{bbs2} nor the quenching of parton jets \cite{jets} is universal 
for different gauge theory plasmas.

It this Letter we propose a lower bound on bulk viscosity $\zeta$ of strongly coupled gauge theories.  Based on holographically
dual computations, we conjecture that a bulk viscosity in a strongly coupled gauge theory plasma in $p$-space dimensions satisfies 
\begin{equation}
\frac{\zeta}{\eta}\ge 2 \left(\frac 1p-c_s^2\right)\,,
\label{bulk}
\end{equation} 
where $c_s$ is the speed of sound. Notice that unlike the shear viscosity bound (\ref{eta1}), our bound (\ref{bulk}) is dynamical: 
as the temperature varies, generically both the speed of sound and the ration of bulk-to-shear viscosities will change. Our claim is that 
the bound  (\ref{bulk}) is correct over all range of temperatures.  

In the following we present evidence in support of the bulk viscosity bound (\ref{bulk}). First, we observe that the bound is saturated 
by the $p+1$ space-time dimensional gauge theory plasma holographically dual to a stack of near-extremal flat Dp-branes \cite{mt}, as well as in
the hydrodynamics of Little String Theory \cite{mt,s}. 
Second, we point out that the bound (\ref{bulk}) remains saturated once above $p$-space dimensional gauge theory is compactified on 
a $k<p$ space-dimensional torus \cite{ss,mt}. Third, we observe that the bound is satisfied (but in general not saturated) in certain $3+1$ strongly 
coupled non-conformal plasma at high temperature \cite{bbs2,cascade}. Finally, we present results \cite{bp} for the bulk viscosity of the 
$\caln=2^*$ gauge theory plasma \cite{n1,n2,n3,n4,n5,bdkl} over a wide range of temperatures, and for various mass deformation parameters.
 We find that the bulk viscosity of the 
$\caln=2^*$ plasma satisfies the bound (\ref{bulk}). As observed in \cite{bdkl}, the $\caln=2^*$ plasma with zero fermion masses undergoes an
interesting phase transition with vanishing speed of sound. A detailed analysis of the critical point \cite{bp} reveals that at the 
transition point the specific heat diverges as $c_V\sim |1-T_c/T|^{-1/2}$. We find that despite the divergent specific heat the bulk viscosity 
at criticality remains finite. We use results for the $\caln=2^*$ gauge theory plasma to estimate the bulk viscosity of QGP at RHIC.

{\it Bulk viscosity of Dp-brane gauge theory plasma.}
$\caln=4$ Yang-Mills plasma at strong coupling is holographically dual to near-extremal stack of D3 branes. In this case conformal 
invariance of the theory implies that 
\begin{equation}
c_s^2=\frac 13\,,\qquad \zeta=0\,.
\label{n4}
\end{equation}
Eq.~(\ref{n4}) was verified in supergravity approximation in \cite{pss1} and beyond the supergravity approximation in 
\cite{cor2}.  Notice that $\caln=4$ plasma trivially satisfies the bound (\ref{bulk}).

In \cite{mt} the authors generalized computation of \cite{pss1} to $p+1$ space-time dimensional gauge theory plasma holographically dual to
near-extremal stack of Dp branes. They found the following dispersion relation for the sound waves  
\begin{equation}
\ww=\sqrt{\frac{5-p}{9-p}}\ \qq -i\ \frac{2}{9-p}\ \qq^2+\cdots\,,
\label{mt1}
\end{equation}
where 
\begin{equation}
\ww\equiv \frac{\omega}{2\pi T}\,, \qquad \qq\equiv \frac{q}{2\pi T}\,.
\end{equation}
Hydrodynamics of a fluid with shear and bulk viscosities $\{\eta,\xi\}$ in $p$-space dimensions predicts the following sound wave dispersion 
\begin{equation}
\omega=c_s\ q- i\ \frac{\eta}{s T}\left(\frac{p-1}{p}+\frac{\zeta}{2\eta}\right)\ q^2+\cdots\,.
\label{sh}
\end{equation}
Using the universality of the shear viscosity (\ref{eta}), one can verify  that the bound (\ref{bulk}) is saturated \cite{mt} in the hydrodynamics of the 
flat Dp branes. It is saturated as well in the hydrodynamics of Little String Theory \cite{mt,s}.

We point out now that the bound (\ref{bulk}) is saturated as well for above strongly coupled gauge theory plasmas compactified on a $k$-dimensional 
torus ( $k<p$ ) \cite{nmt}.  Indeed, upon such a compactification the dispersion relation (\ref{mt1}) will not change --- much like an equation of state it is 
sensitive only to the local properties of the background geometry:
\begin{equation}
\ww_{k<p}=\sqrt{\frac{5-p}{9-p}}\ \qq -i\ \frac{2}{9-p}\ \qq^2+\cdots\,.
\label{mt11}
\end{equation}
On the other hand, the hydrodynamics relation (\ref{sh}) is sensitive to the number of macroscopic ( infinitely extended ) directions:
\begin{equation}
\omega_{k<p}=c_s\ q- i\ \frac{\eta_{k<p}}{s_{k<p} T}\left(\frac{(p-k)-1}{(p-k)}+\frac{\zeta_{k<p}}{2\eta_{k<p}}\right)\ q^2+\cdots\,.
\label{sh1}
\end{equation}  
Again, using the universality of the shear viscosity (\ref{eta}) we find ( see also Eq.~(5.2) of Ref.~\cite{mt} )
\begin{equation}
\frac{\zeta_{k<p}}{\eta_{k<p}}=2\left(\frac{1}{p-k}-c_s^2\right)\,.
\label{nnmt}
\end{equation}
It is precisely for the stated reason the bound (\ref{bulk}) is saturated in Sakai-Sugimoto model in the quenched approximation
\cite{ss}, even though  
\begin{equation}
\frac{\zeta}{\eta}\bigg|_{Sakai-Sugimoto}=\frac{4}{15}\ \ne\ \frac{1}{10}=\frac{\zeta}{\eta}\bigg|_{D4}\,.
\label{compare}
\end{equation}

{\it Bulk viscosity of non-conformal plasma at high temperatures.} A much more nontrivial example is the bulk viscosity of 
non-conformal gauge theory plasma in four dimensions. The computation in the cascading gauge theory \cite{kt,ks} produced \cite{cascade}
\begin{equation}
\frac{\zeta}{\eta}\bigg|_{cascading}=2\left(\frac 13 -c_s^2\right)+{\cal{O}}\left(\left[\frac 13-c_s^2\right]^2\sim \ln^{-2}\frac T\Lambda\right)\,,
\label{casc}
\end{equation}
where $\Lambda$ is the strong coupling scale of the cascading gauge theory.

Likewise, for $\caln=2^*$ gauge theory plasma with bosonic and fermionic mass deformation parameters $m_b\ll T$  and 
$m_f\ll T$, 
\begin{equation}
\frac{\zeta}{\eta}\bigg|_{m_f=0}=\frac{\pi^2\beta_b^\Gamma}{16}\left(\frac 13-c_s^2\right)+{\cal O}\left(\left[\frac 13-c_s^2\right]^2\right)\,,
\label{n2r1}
\end{equation}
where $\beta_b^\Gamma\approx 8.001$ \cite{bbs2}; 
\begin{equation}
\frac{\zeta}{\eta}\bigg|_{m_b=0}=\frac{3\pi\beta_f^\Gamma}{2}\left(\frac 13-c_s^2\right)+{\cal O}\left(\left[\frac 13-c_s^2\right]^2\right)\,,
\label{n2r2}
\end{equation}
where $\beta_f^\Gamma\approx 0.66666$ \cite{mistake}. 

In all cases above we find that the viscosity bound (\ref{bulk}) remains true --- in general, it is no longer saturated. 

\begin{figure}[t]
 \hspace*{-20pt}
\psfrag{cs}{\raisebox{-1ex}{\footnotesize\hspace{0.1cm}$\left(\frac 13-c_s^2\right)$}}
\psfrag{ze}{\raisebox{0ex}{\footnotesize\hspace{0cm}$\frac{\zeta}{\eta}$}}
 \includegraphics[width=3.0in]{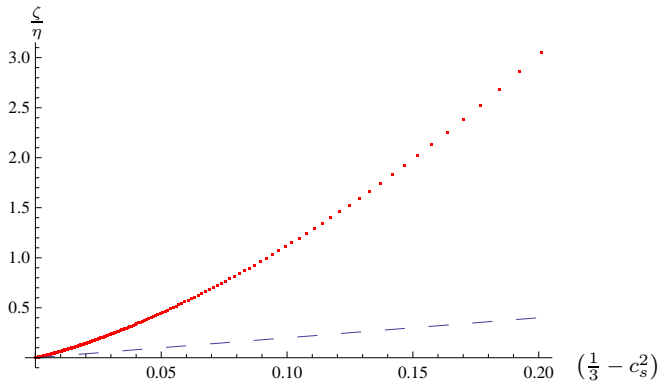}
\caption{Ratio of viscosities  $\frac{\zeta}{\eta}$ versus the speed of sound in 
$\caln=2^*$ gauge theory plasma with zero fermionic mass deformation parameter $m_f=0$. 
The dashed line represents the bulk viscosity bound (\ref{bulk}).}
\label{fig1}
\end{figure}

\begin{figure}[t]
 \hspace*{-20pt}
\psfrag{cs}{\raisebox{-1ex}{\footnotesize\hspace{0.1cm}$c_s^2$}}
\psfrag{ze}{\raisebox{0ex}{\footnotesize\hspace{0cm}$\frac{\zeta}{\eta}$}}
 \includegraphics[width=3.0in]{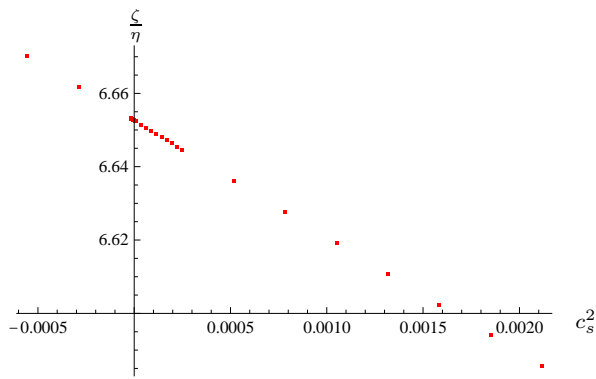}
 \caption{Ratio of viscosities  $\frac{\zeta}{\eta}$ in $\caln=2^*$ gauge theory plasma near the critical point.}
\label{fig2}
\end{figure}

\begin{figure}[t]
 \hspace*{-20pt}
\psfrag{cs}{\raisebox{-1ex}{\footnotesize\hspace{0.1cm}$-\ln\left(\frac{T}{T_c}-1\right)$}}
\psfrag{ze}{\raisebox{0ex}{\footnotesize\hspace{0cm}$\frac{\zeta}{\eta}$}}
 \includegraphics[width=3.0in]{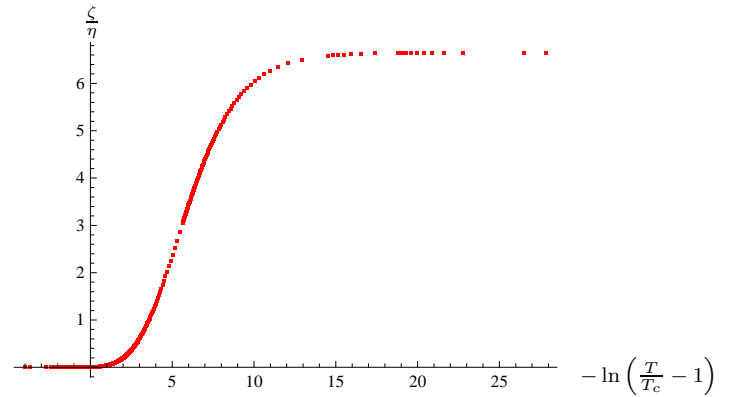}
 \caption{Ratio of viscosities  $\frac{\zeta}{\eta}$ in $\caln=2^*$ gauge theory plasma with zero fermionic mass deformation parameter $m_f=0$. }
\label{fig3}
\end{figure}

\begin{figure}[t]
 \hspace*{-20pt}
\psfrag{cs}{\raisebox{-1ex}{\footnotesize\hspace{0.1cm}$\left(\frac 13-c_s^2\right)$}}
\psfrag{ze}{\raisebox{0ex}{\footnotesize\hspace{0cm}$\frac{\zeta}{\eta}$}}
\psfrag{mtf}{\raisebox{0ex}{\footnotesize\hspace{-0.3cm}$\frac{m}{T}\approx 12$}}
\psfrag{mte}{\raisebox{0ex}{\footnotesize\hspace{0cm}$\frac{m}{T}\to +\infty$}}
 \includegraphics[width=3.0in]{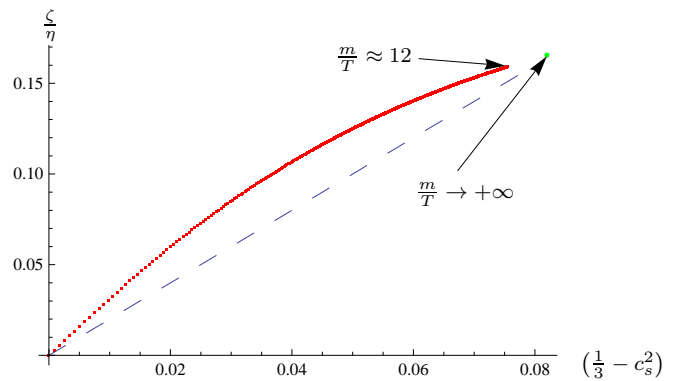}
 \caption{Ratio of viscosities  $\frac{\zeta}{\eta}$ versus the speed of sound in 
$\caln=2^*$ gauge theory plasma with ``supersymmetric'' mass deformation parameters $m_b=m_f=m$. 
The dashed line represents the bulk viscosity bound (\ref{bulk}). We computed the bulk viscosity 
up to $m/T\approx 12$. A single point represents extrapolation of the speed of sound and the 
viscosity ratio to $T\to +0$.}
\label{fig4}
\end{figure}

{\it Bulk viscosity of $\caln=2^*$ plasma.} The strongest support for the bulk viscosity bound 
(\ref{bulk}) comes from study of the $\caln=2^*$ bulk viscosity over the wide range of temperatures.
Such analysis is a  direct extension of the framework presented in \cite{bbs2}. The computations are 
extremely technical and will be detailed elsewhere \cite{bp}. Here, we report only the results of the analysis \cite{request}.

Fig.~\ref{fig1} represents the ratio $\frac{\zeta}{\eta}$ versus the speed of sound in $\caln=2^*$ gauge theory plasma 
with $m_f=0$. This model reaches a critical point with vanishing speed of sound at  $\frac{m_b}{T_c}\approx 2.32591$
\cite{bdkl}. Although near the critical point the specific heat diverges as $c_V\sim |1-T_c/T|^{-1/2}$ \cite{bp} ( also Fig.~8 of \cite{bdkl} ),
we find that the bulk viscosity remains finite,   Fig.~\ref{fig2} and  Fig.~\ref{fig3} .
 
Fig.~\ref{fig4} represents the ratio $\frac{\zeta}{\eta}$ versus the speed of sound in $\caln=2^*$ gauge theory plasma  
 with ``supersymmetric'' mass deformation parameters $m_b=m_f=m$. We did not find any phase transition in this system 
up to temperatures $T \approx \frac{m}{12}$.

The dashed line in Fig.~\ref{fig1} and Fig.~\ref{fig4} represents the bulk viscosity bound 
(\ref{bulk}). In both cases the bound is satisfied.

{\it Estimates for the viscosity of  QGP at RHIC.}
It is tempting to use the $\caln=2^*$ strongly coupled gauge theory plasma results to estimate the 
bulk viscosity of  QGP produced at RHIC. For $c_s^2$ in the range $0.27-0.31$, as in QCD at $T=1.5 T_{deconfinement}$ 
\cite{l1,l2} we find  
\begin{equation}
\frac{\zeta}{\eta}\bigg|_{m_f=0}\approx 0.17-0.61\,,\qquad \frac{\zeta}{\eta}\bigg|_{m_b=m_f=m}\approx 0.07-0.15\,.
\label{qcd}
\end{equation}
Since RHIC produces QGP close to its criticality, we believe that $m_f=0$ $\caln=2^*$ gauge theory model would reflect  physics 
more accurately. If so, it is important to reanalyze the hydrodynamics models of QGP with nonzero bulk viscosity in 
the range given by (\ref{qcd}).

In this Letter we presented some evidence in support of the bulk viscosity bound in strongly coupled gauge theory plasmas. 
It would be interesting to examine other holographic models and test the bound. As in \cite{u2}, it would be interesting to study 
applicability of  the bound in common substances realized in Nature. It appears that common liquids, like water, satisfy the bound \cite{nat1}. 
While the bound is generically satisfied in polyatomic gases \cite{nat2}, it is violated in monoatomic gases \cite{nat3}. 
The bound also appears to be violated  in high-temperature QCD at weak coupling \cite{adm}.
In fact, experimental study of the bulk viscosity in argon at different densities \cite{nat4} demonstrates that its  ratio of bulk-to-shear 
viscosities violates/satisfies the bound at small/large densities.
All this indicates the relevance of the bulk viscosity bound (\ref{bulk}) to strongly coupled 
systems only.

We demonstrated that the bulk viscosity in the $\caln=2^*$ plasma with vanishing fermionic masses has a finite viscosity at the critical point 
with divergent specific heat. The corresponding critical exponent $\alpha=0.5$ ( $c_V\sim |1-T_c/T|^{-\alpha}$ ) coincides with the 
mean-field universal value at the tricritical point \cite{huang}. Such a tricritical point is realized experimentally in solids \cite{sol}.   
It would be interesting to find a fluid with such a universal tricritical point and compare its bulk viscosity with that of the 
$\caln=2^*$ plasma at criticality.

I would like to thank Colin Denniston, Javier Mas,  Rob Myers, Chris Pagnutti, Jim Sethna and Andrei 
Starinets for valuable discussions.
My research at Perimeter Institute is supported in part by the Government
of Canada through NSERC and by the Province of Ontario through MEDT.
I gratefully acknowledges further support by an NSERC Discovery
grant.


\begin{thebibliography}{99}
\bibitem{nmt} A saturation of the bound (\ref{bulk}) upon 
Kaluza-Klein compactification on $k$-dimensional torus was also observed in \cite{mt}.
\bibitem{mistake} There is a mistake in Eq.~(4.37) in \cite{bbs2}: correspondingly to the connection 
coefficient of $dZ^0_\psi/dx$ in Eq.~(4.35), 
the connection coefficient of $dZ^1_\psi/dx$ in Eq.~(4.37) must  be $12 x^2(x^2-1)^2$. Fixing this mistake 
leads to the value of $\beta_f^\Gamma$ presented \cite{bp}.
\bibitem{request} Numerical data is available from the author upon request.
\end{thebibliography}
\end{document}